\documentclass[10pt,a4paper]{article}

\usepackage{scicite}
\usepackage{graphicx}

\usepackage{times}
\usepackage{url}

\newcommand{\nb}{{N\,157B}}
\newcommand{\psr}{PSR\,J0537$-$6910}
\newcommand{\hess}{HESS\,J0537$-$691}

\newcommand{\dorc}{30~Dor~C}
\newcommand{\hessDorc}{HESS\,J0535$-$691}

\newcommand{\nd}{N\,132D}
\newcommand{\hessNd}{HESS\,J0525$-$696}

\newcommand{\sn}{SN\,1987A}
\newcommand{\OtherPsr}{PSR\,J0540$-$6919}

\newcommand{\hessTel}{H.E.S.S.}
\newcommand{\lh}{LH~90}

\newcommand{\HMS}[3]{$#1^{\mathrm{h}}#2^{\mathrm{m}}#3^{\mathrm{s}}$}
\newcommand{\DMS}[3]{$#1^\circ #2' #3''$}

\def\lesssim{\mathrel{\hbox{\rlap{\hbox{\lower3pt\hbox{$\sim$}}}\hbox{\raise2pt\hbox{$<$}}}}}
\def\gtrsim{\mathrel{\hbox{\rlap{\hbox{\lower3pt\hbox{$\sim$}}}\hbox{\raise2pt\hbox{$>$}}}}}

\def\aj{AJ}%
\def\araa{ARA\&A}%
\def\apj{ApJ}%
\def\apjl{ApJ}%
\def\apjs{ApJS}%
\def\apss{Ap\&SS}%
\def\aap{A\&A}%
\def\jcap{J. Cosmology Astropart. Phys.}%
\def\mnras{MNRAS}%
\def\pasp{PASP}%
\def\pasj{PASJ}%
\def\ssr{Space~Sci.~Rev.}%
\def\nat{Nature}%

\def\arcmin{\hbox{$^\prime$}}
\def\arcsec{\hbox{$^{\prime\prime}$}}
\def\utw{\smash{\rlap{\lower5pt\hbox{$\sim$}}}}
\def\udtw{\smash{\rlap{\lower6pt\hbox{$\approx$}}}}

\def\sun{{\lower-2pt\hbox{$_\odot$}}}

\def\g{$\gamma$}

\def\arcmin{\hbox{$^\prime$}}
\def\arcsec{\hbox{$^{\prime\prime}$}}

\newcounter{lastnote}

\twocolumn

\title{The exceptionally powerful TeV \g-ray emitters in the Large Magellanic Cloud}

\author
{
The H.E.S.S. Collaboration
}

\date{}

\begin{document} 
\sloppy

\maketitle

\begin{abstract}{\bf
  The Large Magellanic Cloud, a satellite galaxy of the Milky Way, has
  been observed with the High Energy Stereoscopic System (\hessTel)
  above an energy of 100 billion electron volts for a deep exposure of
  210 hours. Three sources of different types were detected: the
  pulsar wind nebula of the most energetic pulsar known \nb, the
  radio-loud supernova remnant \nd\ and the largest non-thermal X-ray
  shell -- the superbubble \dorc. The unique object SN 1987A is,
  surprisingly, not detected, which constrains the theoretical
  framework of particle acceleration in very young supernova
  remnants. These detections reveal the most energetic tip of a \g-ray
  source population in an external galaxy, and provide via \dorc\ the
  unambiguous detection of \g-ray emission from a superbubble.}
  \end{abstract}

\section*{Introduction}

The origin of cosmic rays (CRs), the very high (VHE, $\gtrsim
10^{11}$~eV), and ultra high ($\gtrsim 10^{18}$~eV) energy particles
that bombard Earth, has puzzled us for over a century. Much progress
has been made during the last decade due to the advent of VHE \g-ray
telescopes. These telescopes detect $\sim 10^{11}-10^{14}$~eV \g-rays
from atomic nuclei (hadronic CRs) collisions with local gas, or from
ultra-relativistic electrons/positrons (leptonic CRs), which produce
\g-ray emission by upscattering low-energy background photons
\cite{HintonHofmann09}. Indeed, a survey of the inner part of the
Milky Way with H.E.S.S., an array of imaging atmospheric Cherenkov telescopes
\cite{HESS_Crab}, revealed a population of supernova remnants (SNRs)
and pulsar wind nebulae (PWNe) emitting \g-rays with energies in
excess of 100 GeV \cite{2005Sci...307.1938A}.

\begin{figure*} 
\includegraphics[width=\textwidth]{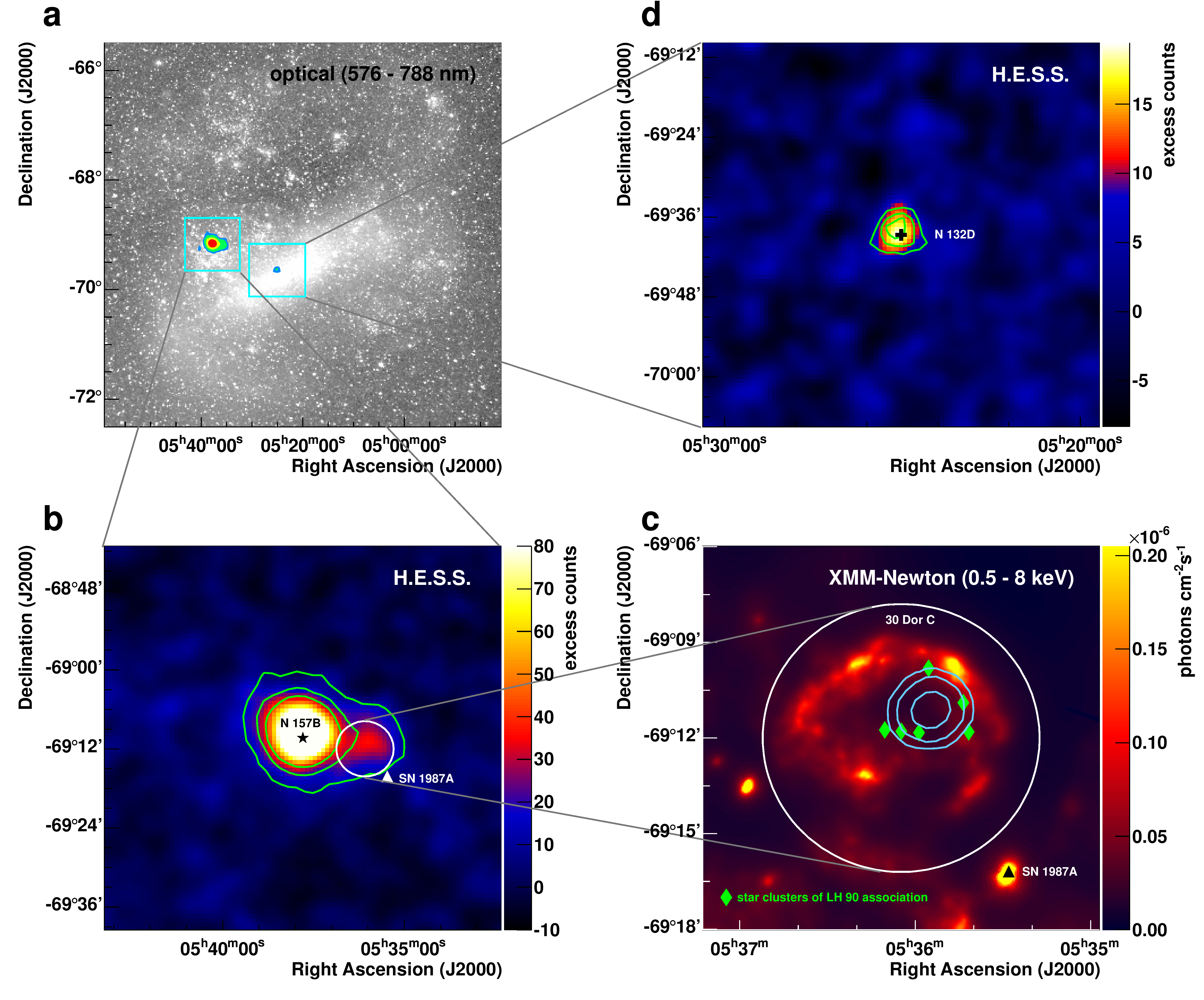}
\caption{
  Sky maps of the Large Magellanic Cloud. 
  a) 
  Optical image of the entire LMC \cite{Mellinger2009}. The boxes denote the regions of interest discussed in this paper. Colours denote levels of 3, 5, 10 and 20\,$\sigma$ statistical significance of the \g-ray signal.
   b) VHE \g-ray emission in the region
  around \nb. The green lines represent contours of 5, 10 and 15\,$\sigma$ statistical significance of the \g-ray signal.  
  c) XMM-Newton
  X-ray flux image of the region of \dorc.
  The superimposed cyan lines represent contours of 68\%, 95\% and 99\% confidence level of the position of the \g-ray source. Diamonds denote the positions of the star clusters of the \lh\ association.
  See supplementary material for details on the X-ray analysis. 
  d) VHE \g-ray emission in the
  region around \nd. The green lines represent contours of
  $3$, $4$ and $5\,\sigma$ statistical significance. 
    The background of the \g-ray emission (in panels b and d) was obtained using the ring background method \cite{BGmodels}. The resulting  
excess sky map is smoothed to the angular resolution of
  the instrument. %
  }
\label{fig:skymap} 
\end{figure*}

Here we report on VHE \g-ray sources detected outside the Milky Way,
namely in the Large Magellanic Cloud (LMC). This satellite galaxy of
the Milky Way has a stellar mass of about  4\% of the Milky Way
\cite{McMillan2011, vanderMarel2006}. Located at a distance of
$\approx$ 50 kpc \cite{2013Natur.495...76P}, it is an irregular galaxy
seen almost face-on \cite{1998ApJ...503..674K}. Consequently, source
confusion is much less of a problem than for the inner Milky Way, and
there is less uncertainty in the distances of the sources. The LMC
stands out among nearby galaxies for its high star formation rate 
per unit mass, which is about a factor of five higher than in the
Milky Way \cite{Harris2009, Robitaille2010}, and contains the best
example of a local starburst, the Tarantula Nebula. The LMC also
harbors numerous massive stellar clusters and SNRs. Among the SNRs is
a unique source, SN1987A, the remnant of the nearest supernova
observed in modern times \cite{1989ARA&A..27..629A}.

High-energy \g-ray emission from the LMC was detected by EGRET
\cite{1992ApJ...400L..67S} and, more recently, by the Fermi Large Area
Telescopes (LAT) \cite{Abdo2010}, which revealed diffuse emission with
an extension of several degrees in diameter, tracing massive
starforming regions.  VHE \g-ray telescopes, like H.E.S.S., besides
providing information on much higher energy CRs, have an angular
resolution of a few arcminutes, which is substantially better
than Fermi-LAT's resolution at \g-ray energies $<10$~GeV.  The
good angular resolution allows H.E.S.S. to identify individual
sources in the LMC.  As we will detail below, a deep
H.E.S.S. observation revealed three luminous sources in the LMC: the
superbubble \dorc, the energetic PWN \nb, and the radio-loud SNR \nd.
Of these sources, only \nb\ was detected previously in a 47-hours
exposure \cite{2012A&A...545L...2H}. The observations extend the
scope of VHE \g-ray astronomy by providing examples of sources from a
population outside the Milky Way. \nb\ and \nd\ belong to known \g-ray
source classes, but both have distinguishing characteristics, \nb\
being powered by the most energetic young pulsar, and \nd\ being one
of the oldest VHE \g-ray emitting SNRs. The superbubble \dorc,
however, provides an unambiguous detection of a superbubble in VHE
\g-rays. Conspicuously absent from our list of three sources is
SN1987A, despite predictions that it should be a bright \g-ray source
\cite{Berezhko2011,Dwarkadas13}.

\section*{H.E.S.S. Observations}
We report on a deep, 210-hours \hessTel\ exposure, targeted at the region of the Tarantula nebula --- corresponding to 30~Doradus (30~Dor) ---
the largest star-forming region in the Local Group of galaxies.
We reconstructed \g-ray showers with an image-fitting analysis
\cite{Mathieu} and cross-checked with a multivariate analysis based on
image parameterization \cite{TMVA,Lu_2013}, with consistent
results. In both analyses, a cut on the uncertainty of the
reconstructed \g-ray direction indicated an angular
resolution of $\approx 0.05^{\circ}$.

Fig.~\ref{fig:skymap}a
shows an optical image of the LMC overlaid with TeV gamma-ray point-source significance contours.
In this dataset, 613 \g\ rays with a statistical significance of $33\,\sigma$ are detected from the PWN \nb.
Figure~\ref{fig:skymap}b 
provides a close-up view of the \g-ray emission from \nb. The diameter of \nb\ of $100\arcsec$ \cite{Chen2006} is of the order of the \hessTel\ angular resolution. Further significant \g-ray emission is detected to the South-West of \nb.

A likelihood fit of a model of two \g-ray sources to the on-source and background sky maps establishes the detection of a second source at an angular distance of $9'$ (corresponding to 130\,pc at a distance of 50\,kpc) from \nb. The model consisting of two sources is preferred by $8.8\,\sigma$  over the model of one single source.
Fig.~\ref{fig:skymap}c 
shows an X-ray image with overlaid contours of confidence of the source position. The position of the second source (RA = $5^{\mathrm{h}} 35^{\mathrm{m}} (55 \pm 5)^{\mathrm{s}}$, Dec =
$- 69^{\circ} 11\arcmin (10 \pm 20)\arcsec$, equinox J2000, $1\,\sigma$ errors) coincides with the superbubble \dorc, the first such source detected in VHE
\g-rays, and thus representing an additional source class in
this energy regime. A \g-ray signal around the energetic pulsar \OtherPsr\ is not detected, despite the presence of an X-ray luminous PWN \cite{2001ApJ...546.1159K}. A flux upper limit (99\% confidence level) is derived at $F_{\gamma}( >1\,\rm{TeV}) < 4.8 \times
10^{-14}\,\rm{ph}\,\rm{cm^{-2}\,s^{-1}}$.

Along with the clear detection of \nb\ and \dorc, evidence for VHE \g-ray emission is observed from the prominent SNR \nd\ (Fig.~\ref{fig:skymap}d).  
The emission peaks at a significance of about $5\,\sigma$ above a background which is estimated from a ring around each sky bin. At the nominal position of the SNR 43 \g\ rays with a statistical significance of $4.7\,\sigma$ are recorded.

\begin{figure*} 
\centering
\includegraphics[width=0.975\textwidth,draft=false]{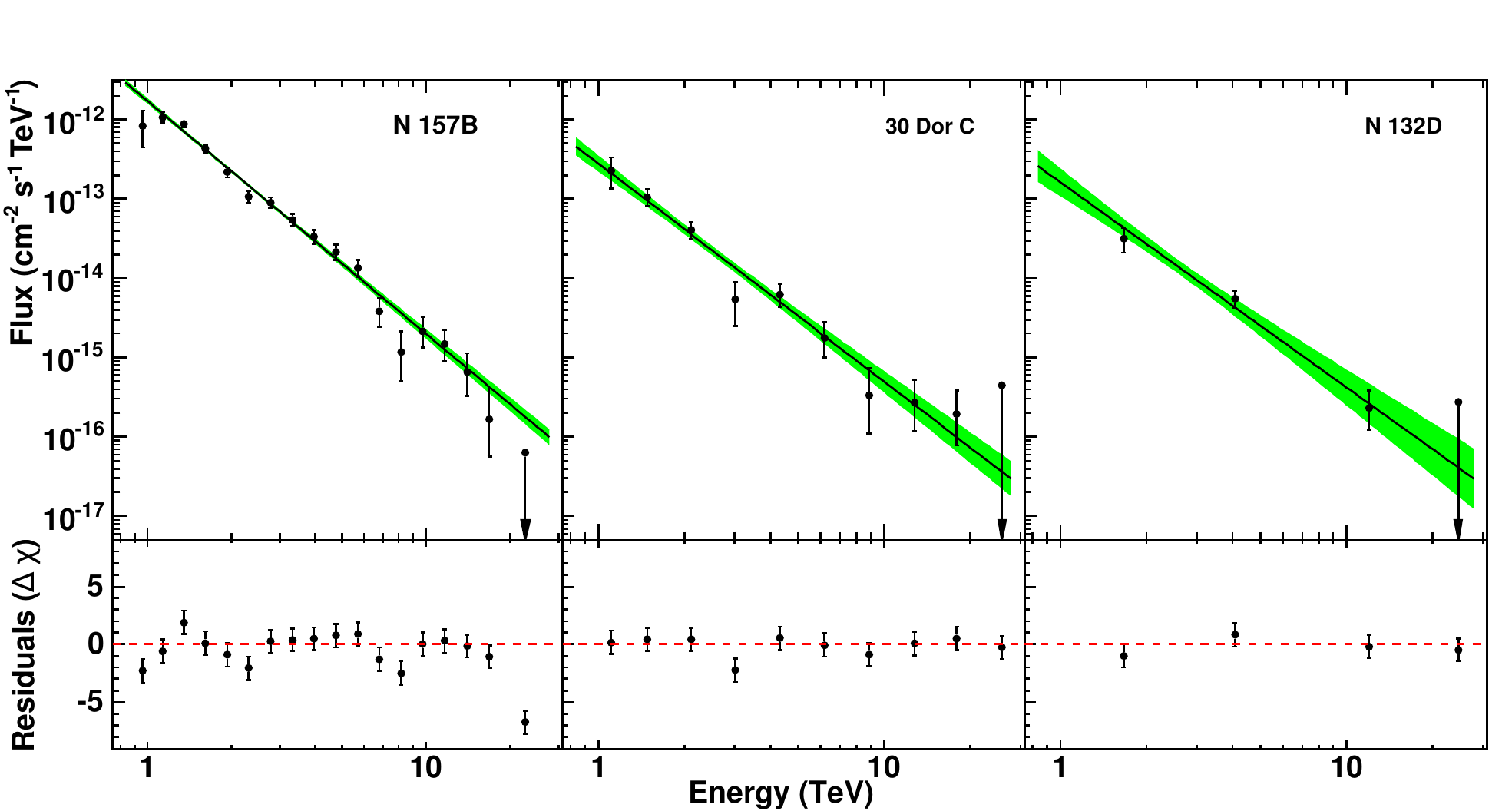}
\caption{
   Gamma ray spectra of \nb, \dorc\ and \nd. Note that the spectral points of \dorc\ are not corrected for 
     spill-over emission from \nb\ (see online supplement). Data points have $1\,\sigma$ error bars, upper limits are at the 99\% confidence level. The bottom panels show the residuals of the data points compared to the best-fit model.}
\label{fig:spectra} 
\end{figure*}

The \g-ray spectra of all three objects are well described by a
power law in energy, $\Phi(E) = d^3N/(dE\,dt\,dA)=\Phi_{0} \left(E/1\,\rm{TeV}\right)^{-\Gamma}$  
(Fig.~\ref{fig:spectra}). 
The best-fit spectral indices and integral
\g-ray luminosities are summarized in Table~\ref{tab:spectral_stat}.

Even with a deep exposure of 210~hours, significant emission from \sn\ is not
detected, and we derive an upper limit on the
integral \g-ray flux of $F_{\gamma}( >1\,\rm{TeV}) < 5.6 \times
10^{-14}\,\rm{ph}\,\rm{cm^{-2}\,s^{-1}}$ at a 99\% confidence level.

\begin{table*}
\centering
\begin{tabular}{lrrr}
\hline
\hline
Source					& \nb				& \dorc				& \nd				\\
H.E.S.S. Identifier		& \hess				& \hessDorc			& \hessNd			\\
\hline
Exposure Time			& 181\,h			& 183\,h			& 148\,h			\\
\g\ rays				& $613$				& $74$				& $43$ 				\\
Significance			& $33.0\,\sigma$	& $8.8\,\sigma$		& $4.7\,\sigma$		\\
\hline
Photon Index $\Gamma$	& $2.8\pm0.1$		& $2.6\pm0.2$		& $2.4\pm0.3$		\\
$\Phi(1\,\rm{TeV})$ $[\rm{10^{-12}\,cm^{-2}\,s^{-1}\,TeV^{-1}}]$
						&  $1.3 \pm 0.1$	& $0.16 \pm 0.04$	& $0.13 \pm 0.05$	\\
$L_{\gamma}(1-10\,\rm{TeV})$ $[\rm{10^{35}\,erg\,s^{-1}}]$
						& $6.8\pm0.3$		& $0.9\pm0.2$		& $0.9\pm0.2$ 		\\
\hline
\hline
\end{tabular}
\caption{Statistics and spectral parameters of the three sources. The exposure time is corrected for the acceptance differences due to different offsets from the camera centre. 
The significances of \nb\ and \nd\ are statistical significances of the \g-ray emission obtained by using formula 17 of \cite{LiMa}. The background was estimated from regions with similar offsets from the camera center as the on-source region.
The significance of \dorc\ is the significance by which a two-source morphology (\nb\ and \dorc) is preferred over a single-source morphology (\nb\ only).
The \g-ray count and the flux of \dorc\ are corrected for spill-over emission from \nb\ (see online supplement). $\Gamma$ is the photon index and $\Phi(1\,\rm{TeV})$ the differential flux at 1\,TeV of a power law fit to the energy spectrum.
The luminosity $L_{\gamma}$ is calculated for an assumed distance of
  50\,kpc \cite{2013Natur.495...76P}. 
  The listed errors are statistical, $1\,\sigma$ errors. 
  Systematic errors are estimated to be $\pm0.3$ for $\Gamma$ and $\pm30\%$ for $\Phi(1\,\rm{TeV})$ \cite{2012A&A...545L...2H}.
  }
\label{tab:spectral_stat} 
\end{table*}

\section*{Discussion of individual sources}

The three VHE emitters belong to different source classes and their
energy output exceeds or at least equals that of their most powerful
relatives in the Milky Way. 

\section*{\dorc}

The superbubble \dorc\ stands out in X-rays as it contains, in the
western part, an X-ray synchrotron-emitting shell with a radius of
47\,pc, which makes it the largest known X-ray synchrotron shell
\cite{Bamba2004,Smith2004,Yamaguchi2009}. X-ray synchrotron emission,
which indicates the presence of VHE electrons, is usually associated
with $100-2000$ year-old SNRs with radii smaller than 25~pc. In
addition, the X-ray synchrotron luminosity of \dorc\ is ten times that
of the archetypal young SNR SN\,1006 \cite{Bamba2004}.  The \dorc\
shell also emits radio and optical radiation \cite{Mathewson1985}, and
appears to have been produced by the stellar winds and supernovae in
the OB association LH~90 (NGC~2044) \cite{LortetTestor1984}.
                                   
The measured H.E.S.S. flux of \dorc\ corresponds to a $1-10$~TeV
\g-ray luminosity of $(0.9 \pm 0.2)\times 10^{35}$\,erg\,s$^{-1}$,
with the best-fit position of the \g-ray emission lying in between the
six identified sub-clusters \cite{LortetTestor1984}. The TeV
  emission can be explained by the production of neutral pions due to
  collisions of hadronic CRs with the background
  plasma. Alternatively, the so-called leptonic emission scenario may
  apply, in which case the TeV emission is the result of Compton
  upscattering of low-energy photons to \g-ray energies, by the same
  population of electrons that is responsible for the X-ray
  synchrotron radiation \cite{HintonHofmann09}.

For the hadronic scenario, a combination of energy in CRs (assumed to
be protons) and density of hydrogen atoms, $n_{\rm H}$, of
$W_{\rm pp}=(0.7-25)\times 10^{52}\,(n_{\rm
  H}/1\rm{cm}^{-3})^{-1}$~erg
is required  (see S1.3). \dorc\ probably experienced $\sim$5
supernova explosions \cite{Smith2004}, which likely provided
$\sim$5$\times 10^{50}$\,erg in CR energy. Hence, the average gas
density should be $n_{\rm H}\gtrsim 20$\,cm$^{-3}$, which is higher
than the density estimate of
$n_{\rm H}\approx 0.1 - 0.4~\mathrm{cm}^{-3}$ based on the X-ray
thermal emission in the southwest \cite{Bamba2004,Kavanagh14}.
However, locations of high densities may be present, if the X-ray
thermal emission comes from smaller radii than the dense outer shell,
or if cool, dense, clumped gas survived inside the otherwise rarified
interior of the bubble \cite{gabici14}.

This hadronic scenario puts constraints on the CR diffusion
coefficient, because the diffusion length scale should be smaller than
the radius of the shell: $l_{\rm diff}= \sqrt{2Dt}\lesssim 47$\,pc
for CRs around 10\,TeV. Therefore,
  $D(10\,{\rm TeV})\lesssim 3.3\times 10^{26} (t/10^6{\rm \
    yr})^{-1}$\,cm$^2$\,s$^{-1}$,
  which, given an age for the superbubble of a few million years, gives
  a much smaller diffusion coefficient than the typical Galactic diffusion
  coefficient of 
  $D(10\,\rm{TeV}) \gtrsim 5\times 10^{29}\,$cm$^2$s$^{-1}$
  \cite{Strong07}.  This small diffusion coefficient requires
magnetic-field amplification combined with turbulent magnetic fields,
as hypothesized by \cite{Bykov2001}.

\begin{figure*} 
  \centering
  \includegraphics[width=0.495\textwidth,draft=false]{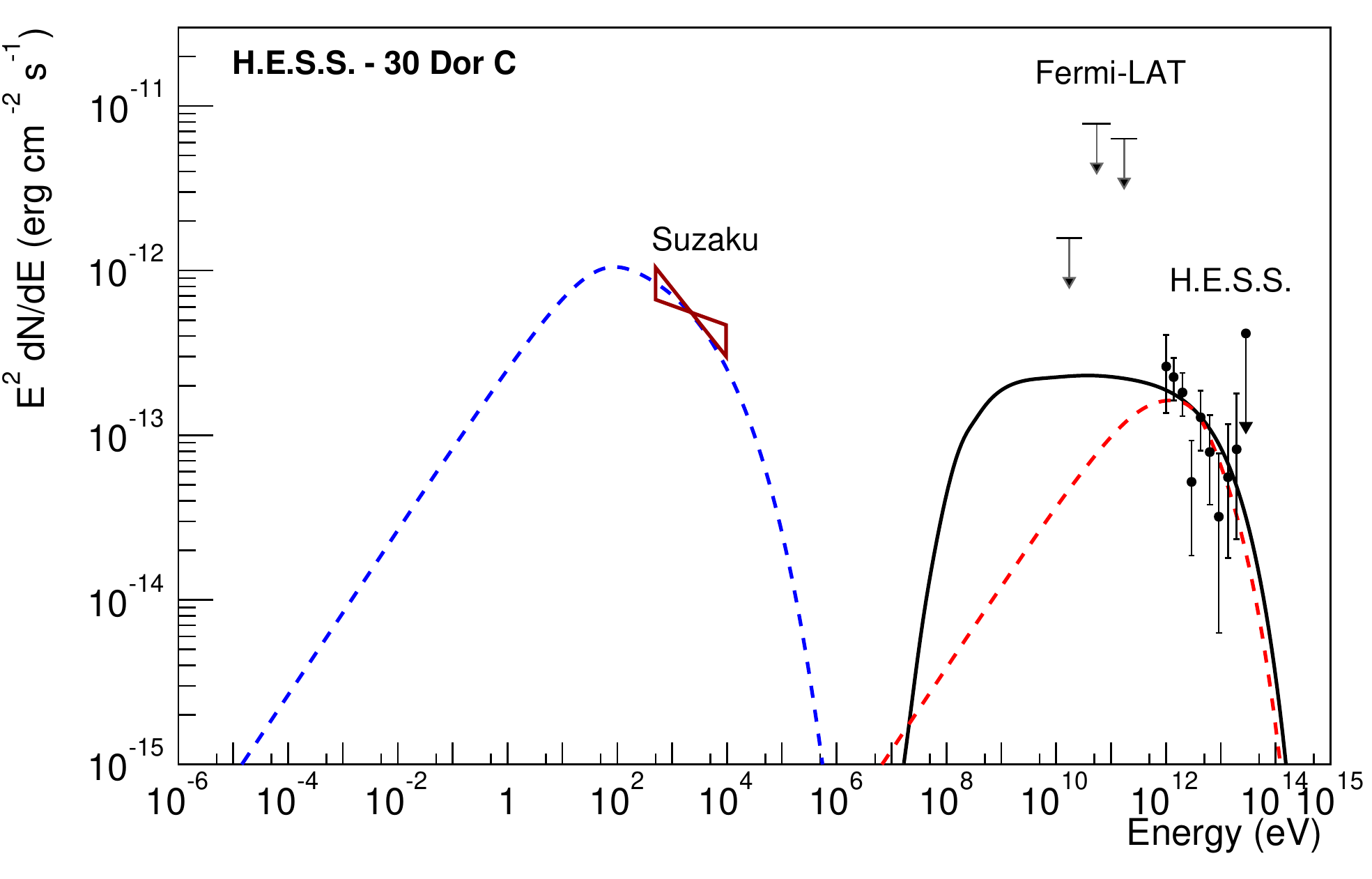}
  \includegraphics[width=0.495\textwidth,draft=false]{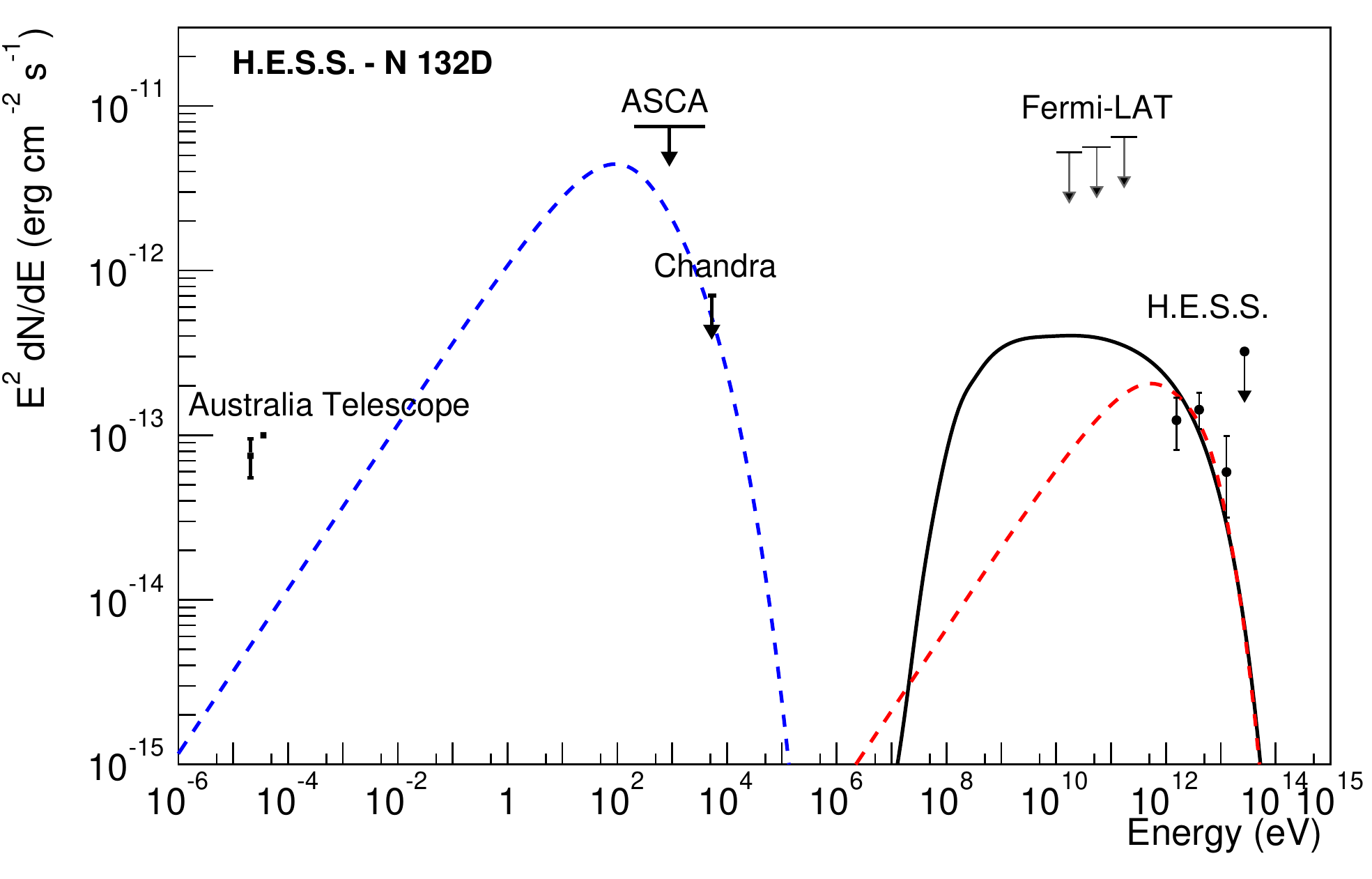}
  \caption{
  Spectral energy distribution of \dorc\ and \nd. The \dorc\
    X-ray data are from \cite{Yamaguchi2009}. For \nd, radio data is
    from \cite{Dickel1995}, and X-ray limits are from
    \cite{Hughes1998} and from re-analysed Chandra data. Both leptonic
    (dashed lines) and hadronic (solid lines) models are shown. 
      For further details on Fermi-LAT data and spectral modelling,
      see S1.2 and S1.3.
      }
  \label{fig:sed_30dorc}
\end{figure*}

X-ray synchrotron emission from \dorc\ requires large shock
velocities, $v_{\rm shock}\gtrsim 3000$\,km\,s$^{-1}$
\cite{aharonian99}. Assuming that this shock originates from an
explosion centered at the superbubble, we obtain a rough estimate of
the age of the western X-ray shell of
$t=0.4 R/v_{\rm shock}\approx 6000$~yr, assuming a Sedov expansion
model ($R=2.8 \times 10^{8}(E t^2/n_\mathrm{H})^{1/5}$~cm). Since the
OB association is much older, this age most likely refers to a
  recent supernova explosion,   whose remnant evolves in the rarified medium of
the superbubble. The Sedov expansion model then gives us a very low
estimate for the density of
$n_{\rm H}\approx 5\times 10^{-4}$\,cm$^{-3}$ for an explosion energy
of $E=10^{51}$\,erg. Although this is very low, it can occur under
certain conditions \cite{Oey04}. This model for the X-ray synchrotron
shell can even be reconciled with the hadronic model of the TeV
emission, if the rarified medium also contains dense clumps. For the
leptonic scenario for the TeV emission, the broad spectral energy
distribution (SED, Fig.~\ref{fig:sed_30dorc})
requires an energy
  in accelerated electrons of $\sim 4\times 10^{48}$\,erg, and average
  magnetic field strength of $15$\,$\mu$G, low compared to most young
  SNRs \cite{Helder12}, but a factor three to four higher than the average magnetic
field in the LMC   \cite{Gaensler2005}.

Although at this stage we cannot rule out either the leptonic or the
hadronic scenario, the H.E.S.S. observations reveal that the
conditions inside the superbubble must be extreme: the hadronic
scenario requires locations with high densities and a high degree of
magnetic turbulence, whereas the leptonic scenario requires the
stellar cluster to be extremely rarified. Moreover, the \g-ray and
X-ray observations suggest active particle acceleration by a very
large, fast expanding shell. This may provide the right conditions for
accelerating some protons to energies exceeding $3\times 10^{15}$~eV,
which is the maximum energy detected for Galactic CRs.  These
  observations, therefore, lend support to the view expressed in
  \cite{Bykov2001,Parizot2004,Ferrand2010} that superbubbles may
  provide the right conditions for particle acceleration to very high
  energies, because they are thought to contain very turbulent
  magnetic fields and they are large enough to contain VHE particles
  for up to millions of year.

In the Milky Way, the most closely related object to \dorc\  is
the stellar cluster Westerlund 1 \cite{Abramowski2012}, which,
however, has a completely different X-ray morphology. More
importantly, it is not clear whether the \g-rays originate from the
cluster wind itself, a PWN or from the numerous supernovae that
exploded inside Westerlund 1 in the recent past. Since a large
fraction of supernovae are thought to go off in superbubbles, this
first unambiguous detection of VHE \g-rays from a superbubble may have
broad implications for the circumstances in which a large fraction of
CRs are accelerated.

\section*{\nb}

The source \hess\ is coincident with the PWN \nb, which surrounds the
pulsar \psr. PWNe are nebulae of ultra-relativistic particles driven
by highly-magnetized, fast-rotating neutron stars that convert a
considerable amount of their spin-down energy into a particle
wind. The archetypal Crab nebula is one of the brightest sources of
non-thermal radiation in the sky and powered by the pulsar with the
highest spin-down energy known in the Milky Way
\cite{Buehler2013}. With comparably extreme rotational energy loss
rates, \nb\ ($\dot{E}=4.9\times10^{38}$\,erg\,s$^{-1}$) and the Crab
nebula ($\dot{E}=4.6\times10^{38}$\,erg\,s$^{-1}$) appear to be
twins. The study of \nb\ thus provides the unique opportunity to
compare two extreme PWNe, and to disentangle object-specific and
generic properties.

\begin{figure*} 
\centering
\includegraphics[width=0.75\textwidth,draft=false]{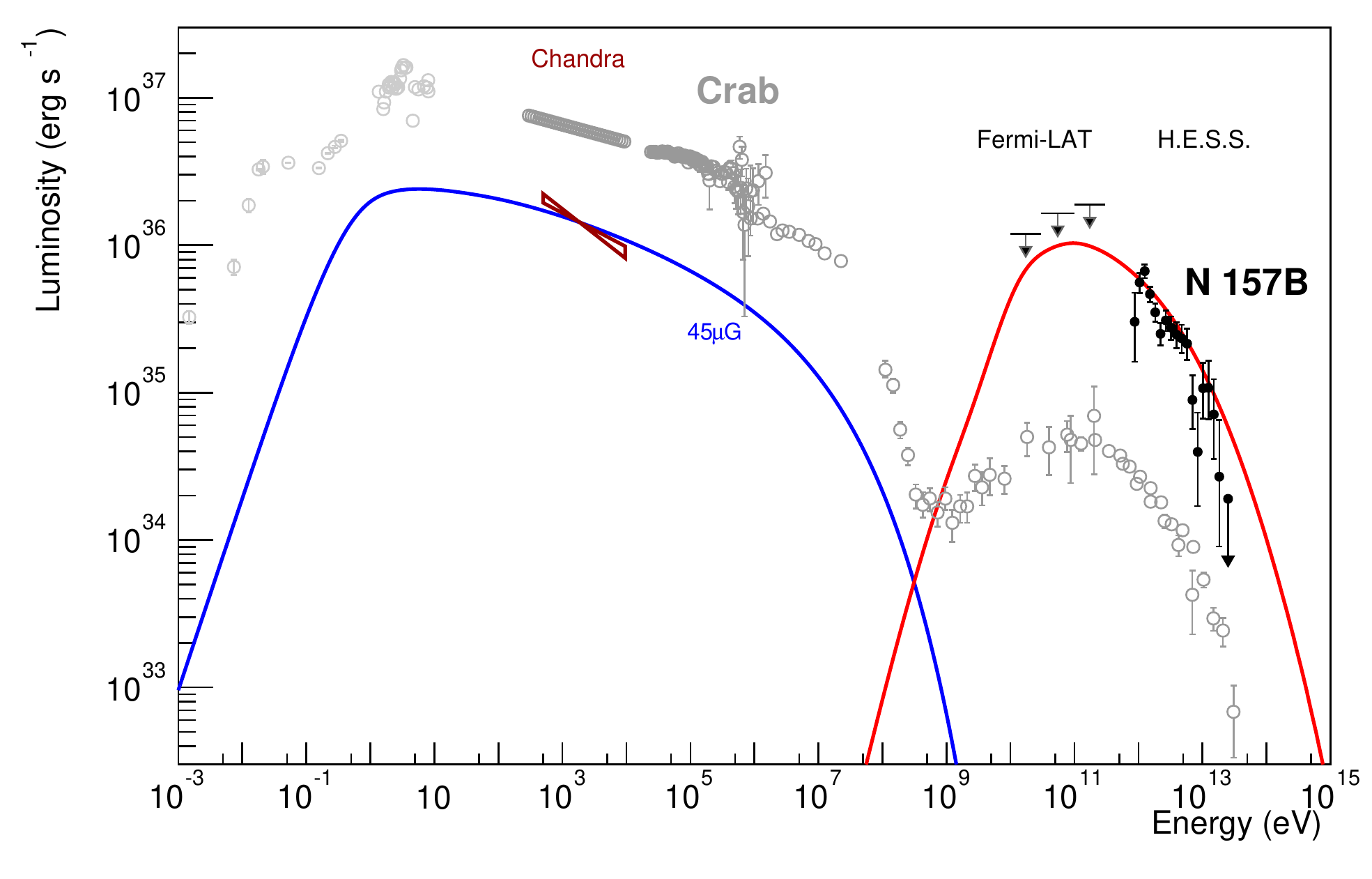}
\caption{
Intrinsic SED of \nb\ (black) and the Crab nebula (grey). The
  model shown assumes the same injection parameters as derived for the
  Crab nebula ($E_{\rm min}$ = 400\,GeV, $E_c$ = 3.5\,PeV,
  $\Gamma_e = 2.35$ \cite{Bucciantini2011}). The magnetic field
  required to explain the \emph{Chandra} data of \nb\ in the highest
  possible radiation fields is 45$\mu$G. A significantly better fit to
  the \emph{Chandra} data is obtained with $\Gamma_{e}=2.0$, but
  requires a much lower cut-off of
  $E_{\mathrm{c}} \lesssim 100$\,TeV.
  }
\label{fig:sed_comp}
\end{figure*}

Given a population of ultra-relativistic electrons and positrons
forming the PWN, the X-ray luminosity is determined by the strength of
the magnetic field and the \g-ray luminosity by the intensity of
radiation fields which serve as targets for the inverse Compton
upscattering. If the radiation fields are known, the magnetic field
can be inferred from the combination of X-ray and \g-ray
measurements. \nb\ is likely associated with the LH~99 star cluster
\cite{Chu1992, Wang1998, Chen2006}, and therefore embedded in strong
infrared radiation fields (see S1.5). In this environment, the
magnetic field in the PWN must be rather weak, not larger than
45\,$\mu$G, in order to explain the multiwavelength data
(Fig.~\ref{fig:sed_comp}). 
When considering the region from which the
hard X-ray emission is coming, the total energy in the magnetic field
is $W_{B,\rm tot} = 1.4\times10^{47}$\,erg -- an order of magnitude
smaller than the energy in $>$$400$\,GeV
electrons. The derived maximum magnetic field is also much lower than
that inferred for the Crab nebula
($\sim$$124\,\mu$G
\cite{Meyer2010}), and suggests at least a factor $\sim$$7$
lower magnetic pressure. As most of the electrons that radiate in the
\emph{Chandra}, X-ray and H.E.S.S., TeV domains have very short
lifetimes ($\le$300
years), the energy in ultra-relativistic particles in \nb\ can be
inferred independently of the spin-down evolution of the pulsar. For
the model shown in Fig.~\ref{fig:sed_comp}, 
a constant fraction of $11\%$
of the current spin-down power of \nb\ needs to be injected into the
nebula in the form of relativistic electrons (compared with 50\% for
the Crab nebula under the same model). This fraction converted into
X-ray and TeV emission is rather insensitive to the spectral index of
injected electrons and the spin-evolution or braking index of the
pulsar and only relies on the association of \nb\ with LH~99 (see
  S1.5 for more information).

In this high-radiation field scenario, the situation for the
Crab nebula is very different from \nb. Not only is the best-fit
electron spectrum of \nb\ harder $(\Gamma_e
= 2.0$ vs. 2.35), exhibiting a lower cut-off energy $(E_c =
100$\,TeV vs. 3.5\,PeV), but much of the spin-down energy of \nb\ is
also hidden and is not carried by ultra-relativistic particles or
magnetic fields. The remainder of the available rotational energy is
likely to be fed into electrons with energies $\leq$400\,GeV
that radiate at lower photon energies, adiabatic expansion, and/or
particles escaping into the interstellar medium via diffusive escape
(e.g., \cite{Hinton2011}). It therefore appears that \nb\ is such a
bright \g-ray emitter because of the enhanced radiation fields,
despite the fact that it is apparently a much less efficient particle
accelerator than the Crab nebula.

\section*{\nd}
In addition to the two unambiguously detected sources, we find strong
evidence for a third source at the position of the core-collapse SNR
\nd. \nd\ is a SNR with strong thermal X-ray emission, which has been
used to estimate a pre-shock density of
$n_\mathrm{H}\approx 2.6~\mathrm{cm}^{-3}$ \cite{Hughes1998}, a high
explosion energy of $\sim 6\times 10^{51}$~erg \cite{Hughes1998}, and
an age of $\sim 6000$~yr, based on a Sedov model.  Such X-ray bright
SNRs are predicted to be \g-ray emitters \cite{KatzWaxman2008}. \nd\
is also luminous in the radio \cite{Dickel1995} and infrared bands
\cite{Tappe2006}. \nd\ is often compared to the brightest radio source
Cas A, which, like \nd, is an oxygen-rich SNR. \nd\ has a higher
infrared luminosity \cite{Tappe2006}, but its radio luminosity is 50\%
that of Cas A. This is still remarkable given that \nd\ has a
kinematic age of $\sim 2500$~yr \cite{Vogt2011}, whereas Cas A is
$\sim 330$~yr old and declines in luminosity by about 0.8\% per year.
The radio properties have been used to infer a magnetic field strength
of $\sim 40~\mathrm{\mu G}$ \cite{Dickel1995}. The discrepancy between
the age estimate based on the X-ray emission and the kinematic age may
indicate that the supernova exploded within a bubble created by the
progenitor star's wind before encountering the high density material
it now interacts with.

The \g-ray flux measured by H.E.S.S. translates to a $1-10$~TeV \g
-ray luminosity of
$(0.9\pm0.2) \times 10^{35}(d/50\mathrm{kpc})^{2}\,\mathrm{erg/s}$.
Assuming that the \g-ray emission is caused by neutral-pion
production, this luminosity implies an energy of
$10^{52} (n_{\rm H}/1\mathrm{cm}^{-3})^{-1}
(d/50\mathrm{kpc})^{-2}\,\mathrm{erg}$
in relativistic protons.  A hadronic origin of the \g-ray emission,
therefore, implies either a large CR-energy fraction of 17\% of the
explosion energy, for an estimated post-shock density of
$n_\mathrm{H}\approx 10~\mathrm{cm}^{-3}$ \cite{Hughes1998}, or the
gas density is higher than the x-ray-based estimates. The latter is
plausible given that \nd\ appears to interact with dense, shocked
interstellar clouds, seen in the optical and the infrared bands
\cite{Tappe2006}. It is interesting to compare \nd\ to the most
luminous Galactic SNR detected at TeV energies, HESS~J1640$-$465: both
SNRs are believed to interact with a wind-blown cavity wall, to
possibly have similar ages and sizes \cite{Abramowski2014,Vogt2011},
and to have transferred a large fraction of their explosion energies
into CRs.

The bright radio synchrotron luminosity of \nd\ and the tentative
claim of X-ray synchrotron emission from this source \cite{Xiao2008}
also raises the possibility that the \g-ray emission is caused by
inverse Compton scattering of low-energy photons. In and around \nd\
the radiation energy density is dominated by the bright infrared flux
from dust inside the SNR, and can be roughly estimated to be at least
$u_\mathrm{rad}\approx 1.0~\mathrm{eV}\,\mathrm{cm}^{-3}$. This
leptonic scenario requires that the average magnetic-field strength
needs to be $\sim 20~\mu$G, somewhat lower, but still consistent with
the equipartition value (see S1.3). However, this leptonic scenario
critically depends on whether the 4-6 keV X-ray continuum emission
indeed contains a significant synchrotron component.

Whatever the emission mechanism for the \g-ray emission from \nd, it
is an exciting new \g-ray-emitting SNR, because its age lies in the
gap between young ($<2000$\,yr) TeV-emitting SNRs, and old
($\gtrsim 10000$\,yr) TeV-quiet SNRs. The latter can be bright
pion-decay sources, but their spectra appear to be cut off above
$\sim 10$ GeV. \nd\ provides, therefore, an indication of how long
SNRs contain CRs with energies in excess of $10^{13}$~eV.

\section*{\sn}

\sn, the only naked-eye SN event since the Kepler SN (AD 1604), 
has been extensively observed at all wavelengths from the radio to the soft 
\g-ray band, providing invaluable insights  
into the evolution of a core-collapse SNR in its early stage \cite{Immler2007}.

It has been suggested that even in the early stages of the SNR development,
the shock wave, which is heating the dense circumstellar medium (CSM) structured by stellar winds of the
progenitor star, should have led
to efficient acceleration of VHE nuclear CRs, accompanied by strong magnetic field amplification 
through CR-induced instabilities \cite{Bell13,Berezhko2011}.
In collisions of the CRs with CSM particles, \g-rays are produced. 
Estimates for the \g-ray flux \cite{Berezhko2011,Dwarkadas13}
strongly depend on the magnetic field topology
and on the properties of the non-uniform
CSM \cite{Chevalier1995}, making flux estimates uncertain by at least a factor
of 2 \cite{Berezhko2011}.

Based on a nonlinear kinetic theory of CR acceleration, successfully 
applied to several young Galactic SNRs, the volume-integrated \g-ray 
flux at TeV energies, $F_{\gamma}(>1\,\rm{TeV})$, from \sn\ was predicted to 
be rising in time, and to have reached a level of
$\approx 2.5 \times 10^{-13}$\,ph\,cm$^{-2}$\,s$^{-1}$ in the year 2010 
\cite{Berezhko2011}. An  analysis with
different assumptions on CSM properties and a more phenomenological
approach to CR acceleration resulted in a predicted flux of 
$\sim 8 \times 10^{-14}$
ph\,cm$^{-2}$\,s$^{-1}$ in the year 2013 \cite{Dwarkadas13}. The H.E.S.S. upper limit
$F_{\gamma}(> 1\,\rm{TeV}) < 5 \times 10^{-14}$\,ph\,cm$^{-2}$\,s$^{-1}$ 
at a $99\%$ confidence level,
obtained from observations made between 2003 and 2012, being below 
the aforementioned predictions and a factor 
of 3 below similar estimates for the year 2005, therefore places
constraints on the models despite their uncertainties.

The H.E.S.S. upper limit on the \g-ray flux translates into an upper limit for the  
\g-ray luminosity 
of $L_{\gamma}(>1\,\rm{TeV}) < 2.2\times 10^{34}$\,erg/s, which can 
be used to derive 
an approximate upper limit on the energy of the accelerated particles, $W_{pp}$, 
for a given average target density. 
Multi-wavelength studies of \sn\ suggest that 
the shock at the current epoch has reached and is interacting with the so-called equatorial ring,
for which gas densities ranging from $10^3$ cm$^{-3}$ to $3\times10^4$ cm$^{-3}$ 
have been found \cite{Mattila2010}. Thus one finds a conservative upper limit,
 $W_{pp} \lesssim 1.4 \times 10^{48}f^{-1}$ erg, where $0<f<1$ is the fraction of 
accelerated particles that are interacting with the dense regions. 
This upper limit on the energy of accelerated CR particles corresponds
to $0.15f^{-1}\,\%$ of  the explosion energy of $10^{51}$ erg.

Assuming a spherically-symmetric distribution of accelerated particles, one can estimate $f\sim0.2$ with the geometry of the equatorial ring found in \cite{Ng2011}.   
This translates to $W_{pp} \lesssim 9 \times 10^{48}$ erg, implying that 
less than $1\%$ of the explosion energy is carried by
accelerated CR nuclei. 
This fraction is rather small compared to typical values
 of $\sim10\%$ for young SNRs (of ages $\sim 1000-2000$~years),
 but is not unreasonable for a very young object like \sn. 
 
\section*{Summary}

With the deep H.E.S.S. observations of the LMC, we have detected three
luminous examples of CR sources in an external galaxy. These sources
detected in \g\ rays include a superbubble and counterparts to the
most luminous sources in the Milky Way. \nb\ provides a counterpart to
the Crab Nebula, but its electron acceleration efficiency is five
times less than for the Crab nebula, and its magnetic field pressure
is seven times less. \nd\ has been long regarded an older version of
the brightest Galactic radio SNR Cas A, and is one of the most
radio-luminous SNRs known.  \nd\ is also remarkable in that it is one
of the oldest VHE \g-ray emitting SNRs.  With the three detected
sources, we increase our understanding of the variety of VHE \g-ray
sources, which will likely require observations with the future
Cherenkov Telescope Array \cite{CTA}, which should be an order of
magnitude more sensitive than H.E.S.S.

\section*{Acknowledgements}
The support of the Namibian authorities and of the University of
Namibia in facilitating the construction and operation of H.E.S.S. is
gratefully acknowledged, as is the support by the German Ministry for
Education and Research (BMBF), the Max Planck Society, the German
Research Foundation (DFG), the French Ministry for Research, the
CNRS-IN2P3 and the Astroparticle Interdisciplinary Programme of the
CNRS, the U.K. Science and Technology Facilities Council (STFC), the
IPNP of the Charles University, the Czech Science Foundation, the
Polish Ministry of Science and Higher Education, the South African
Department of Science and Technology and National Research Foundation,
and by the University of Namibia. We appreciate the excellent work of
the technical support staff in Berlin, Durham, Hamburg, Heidelberg,
Palaiseau, Paris, Saclay, and in Namibia in the construction and
operation of the equipment. The H.E.S.S. collaboration will make public 
the smoothed excess sky map and the associated correlated significance map as shown in Figure 1, 
together with the source spectral points, on the HESS website on
the link to this publication: 
\url{http://www.mpi-hd.mpg.de/hfm/HESS/pages/publications}.
We would like to thank Boaz Katz, Eli Waxman and Ranny Budnik for
their external proposal supporting observations of the SNR N 132D
based on their work on \g-ray emission from shell-type SNRs
\cite{KatzWaxman2008}.

\section*{The H.E.S.S. Collaboration}

A.~Abramowski $^{1}$, F.~Aharonian $^{2,3,4}$, F.~Ait Benkhali $^{2}$, A.G.~Akhperjanian $^{5,4}$, E.O.~Ang\"uner $^{6}$,, M.~Backes $^{7}$,
 S.~Balenderan $^{8}$, A.~Balzer $^{9}$, A.~Barnacka $^{10,11}$, Y.~Becherini $^{12}$
% ...
\\
\\
$^{1}$ Universit\"at Hamburg, Institut f\"ur Experimentalphysik, Luruper Chaussee 149, D 22761 Hamburg, Germany, 
$^{2}$ Max-Planck-Institut f\"ur Kernphysik, P.O. Box 103980, D 69029 Heidelberg, Germany, 
$^{3}$ Dublin Institute for Advanced Studies, 31 Fitzwilliam Place, Dublin 2, Ireland, 
$^{4}$ National Academy of Sciences of the Republic of Armenia,  Marshall Baghramian Avenue, 24, 0019 Yerevan, Republic of Armenia , 
$^{5}$ Yerevan Physics Institute, 2 Alikhanian Brothers St., 375036 Yerevan, Armenia, 
$^{6}$ Institut f\"ur Physik, Humboldt-Universit\"at zu Berlin, Newtonstr. 15, D 12489 Berlin, Germany, 
$^{7}$ University of Namibia, Department of Physics, Private Bag 13301, Windhoek, Namibia, 
$^{8}$ University of Durham, Department of Physics, South Road, Durham DH1 3LE, U.K., 
$^{9}$ GRAPPA, Anton Pannekoek Institute for Astronomy, University of Amsterdam,  Science Park 904, 1098 XH Amsterdam, The Netherlands, 
$^{10}$ Obserwatorium Astronomiczne, Uniwersytet Jagiello{\'n}ski, ul. Orla 171, 30-244 Krak{\'o}w, Poland, 
$^{11}$ now at Harvard-Smithsonian Center for Astrophysics,  60 Garden St, MS-20, Cambridge, MA 02138, USA, 
$^{12}$ Department of Physics and Electrical Engineering, Linnaeus University,  351 95 V\"axj\"o, Sweden, 
$^{13}$ Institut f\"ur Theoretische Physik, Lehrstuhl IV: Weltraum und Astrophysik, Ruhr-Universit\"at Bochum, D 44780 Bochum, Germany, 
$^{14}$ GRAPPA, Anton Pannekoek Institute for Astronomy and Institute of High-Energy Physics, University of Amsterdam,  Science Park 904, 1098 XH Amsterdam, The Netherlands, 
$^{15}$ Institut f\"ur Astro- und Teilchenphysik, Leopold-Franzens-Universit\"at Innsbruck, A-6020 Innsbruck, Austria, 
$^{16}$ Laboratoire Leprince-Ringuet, Ecole Polytechnique, CNRS/IN2P3, F-91128 Palaiseau, France, 
$^{17}$ now at Santa Cruz Institute for Particle Physics, Department of Physics, University of California at Santa Cruz,  Santa Cruz, CA 95064, USA, 
$^{18}$ Centre for Space Research, North-West University, Potchefstroom 2520, South Africa, 
$^{19}$ LUTH, Observatoire de Paris, CNRS, Universit\'e Paris Diderot, 5 Place Jules Janssen, 92190 Meudon, France, 
$^{20}$ LPNHE, Universit\'e Pierre et Marie Curie Paris 6, Universit\'e Denis Diderot Paris 7, CNRS/IN2P3, 4 Place Jussieu, F-75252, Paris Cedex 5, France, 
$^{21}$ Institut f\"ur Astronomie und Astrophysik, Universit\"at T\"ubingen, Sand 1, D 72076 T\"ubingen, Germany, 
$^{22}$ Laboratoire Univers et Particules de Montpellier, Universit\'e Montpellier 2, CNRS/IN2P3,  CC 72, Place Eug\`ene Bataillon, F-34095 Montpellier Cedex 5, France, 
$^{23}$ DSM/Irfu, CEA Saclay, F-91191 Gif-Sur-Yvette Cedex, France, 
$^{24}$ Astronomical Observatory, The University of Warsaw, Al. Ujazdowskie 4, 00-478 Warsaw, Poland, 
$^{25}$ Instytut Fizyki J\c{a}drowej PAN, ul. Radzikowskiego 152, 31-342 Krak{\'o}w, Poland, 
$^{26}$ School of Physics, University of the Witwatersrand, 1 Jan Smuts Avenue, Braamfontein, Johannesburg, 2050 South Africa, 
$^{27}$ Landessternwarte, Universit\"at Heidelberg, K\"onigstuhl, D 69117 Heidelberg, Germany, 
$^{28}$ Oskar Klein Centre, Department of Physics, Stockholm University, Albanova University Center, SE-10691 Stockholm, Sweden, 
$^{29}$ Wallenberg Academy Fellow, , 
$^{30}$  Universit\'e Bordeaux 1, CNRS/IN2P3, Centre d'\'Etudes Nucl\'eaires de Borde\author
{
H.E.S.S. Collaborationaux Gradignan, 33175 Gradignan, France, 
$^{31}$ Funded by contract ERC-StG-259391 from the European Community, , 
$^{32}$ School of Chemistry \& Physics, University of Adelaide, Adelaide 5005, Australia, 
$^{33}$ APC, AstroParticule et Cosmologie, Universit\'{e} Paris Diderot, CNRS/IN2P3, CEA/Irfu, Observatoire de Paris, Sorbonne Paris Cit\'{e}, 10, rue Alice Domon et L\'{e}onie Duquet, 75205 Paris Cedex 13, France, 
$^{34}$ Univ. Grenoble Alpes, IPAG,  F-38000 Grenoble, France; CNRS, IPAG, F-38000 Grenoble, France, 
$^{35}$ Department of Physics and Astronomy, The University of Leicester, University Road, Leicester, LE1 7RH, United Kingdom, 
$^{36}$ Nicolaus Copernicus Astronomical Center, ul. Bartycka 18, 00-716 Warsaw, Poland, 
$^{37}$ Institut f\"ur Physik und Astronomie, Universit\"at Potsdam,  Karl-Liebknecht-Strasse 24/25, D 14476 Potsdam, Germany, 
$^{38}$ Laboratoire d'Annecy-le-Vieux de Physique des Particules, Universit\'{e} de Savoie, CNRS/IN2P3, F-74941 Annecy-le-Vieux, France, 
$^{39}$ DESY, D-15738 Zeuthen, Germany, 
$^{40}$ Universit\"at Erlangen-N\"urnberg, Physikalisches Institut, Erwin-Rommel-Str. 1, D 91058 Erlangen, Germany, 
$^{41}$ Centre for Astronomy, Faculty of Physics, Astronomy and Informatics, Nicolaus Copernicus University,  Grudziadzka 5, 87-100 Torun, Poland, 
$^{42}$ Department of Physics, University of the Free State,  PO Box 339, Bloemfontein 9300, South Africa, 
$^{43}$ Charles University, Faculty of Mathematics and Physics, Institute of Particle and Nuclear Physics, V Hole\v{s}ovi\v{c}k\'{a}ch 2, 180 00 Prague 8, Czech Republic, 
$^{44}$ GRAPPA, Institute of High-Energy Physics, University of Amsterdam,  Science Park 904, 1098 XH Amsterdam, The Netherlands}
\\
\normalsize{
$^\ast$To whom correspondence should be addressed; 
E-mail: nukri.komin@wits.ac.za, chia-chun.lu@mpi-hd.mpg.de,
michael.mayer@physik.hu-berlin.de, stefan.ohm@desy.de, j.vink@uva.nl}

\clearpage
\vfill
\appendix

\centerline{\LARGE\bf Supplementary Materials}
\section{Materials and Methods}

\subsection{Correction factor for \dorc\ flux}
The superbubble \dorc\ and the PWN \nb\ have an angular distance of
$\approx 0.15^{\circ}$, while the angular resolution ($68\%$
containment radius of the point-spread-function (PSF)) of H.E.S.S. is
$\approx 0.05^{\circ}$.  The flux of \g-ray sources is calculated from
the excess of \g-like events in a pre-defined on-region. As
the reconstructed flux of \dorc\ has been estimated within a test
region of $0.07^{\circ}$ around the center of the X-ray superbubble, a
part of the \g-rays from \nb\ lies in this on-region due to the tails
of the PSF. These additional \g-rays will result in a larger
reconstructed flux for \dorc, and need to be corrected for in order to
estimate the ``true'' flux of \dorc.

\begin{figure} 
\begin{center}
   \includegraphics[width=0.5\textwidth,draft=false]{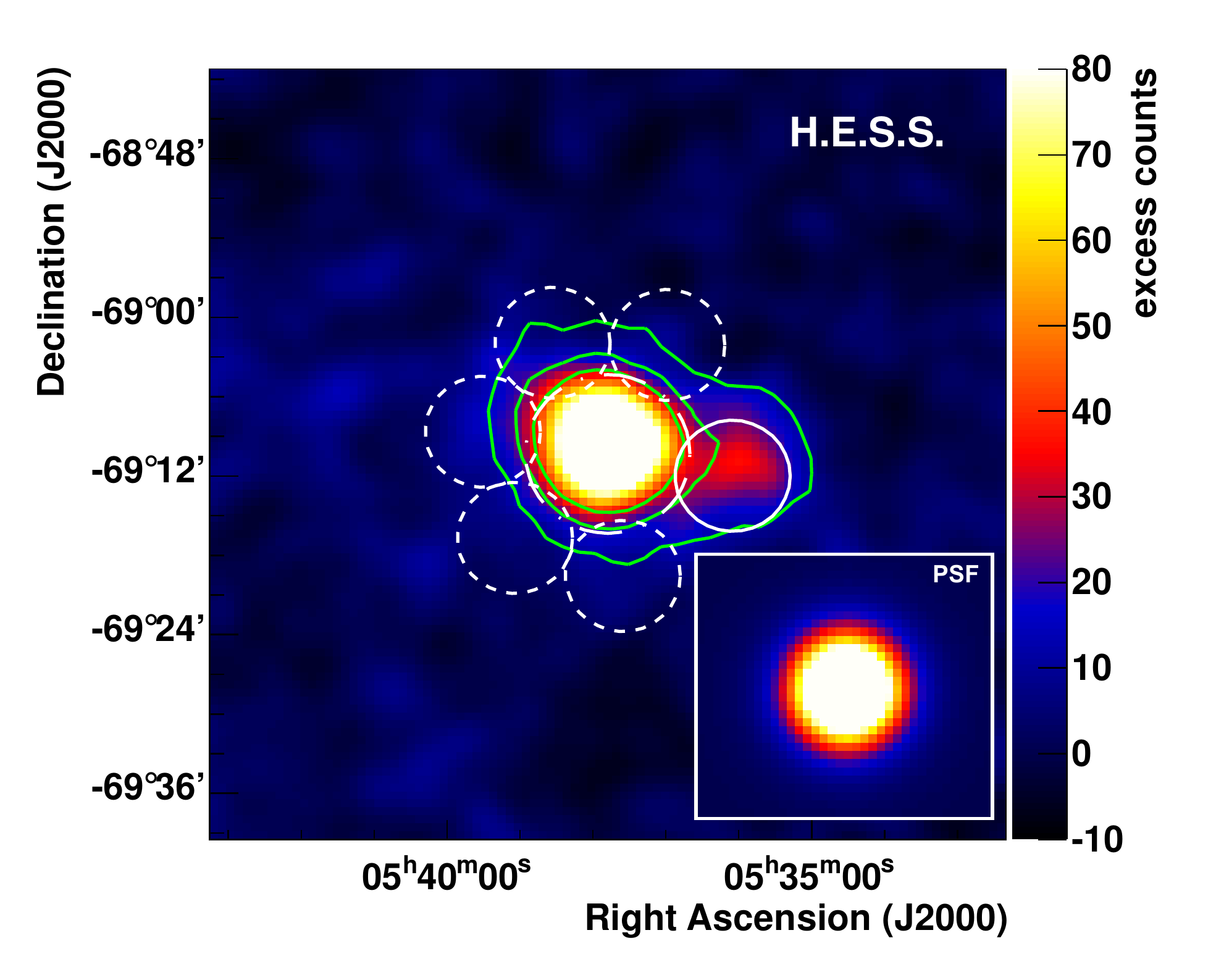}
   \caption{On (solid circle) and test (dashed circles) regions to
     estimate the flux from \dorc\ on top of the smoothed \g-ray
     excess map. The dash-dotted circle denotes the region of interest
     of \nb. The inset shows the simulation of a point-like source.}
\label{fig:test_regions} 
\end{center}
\end{figure}

The H.E.S.S. PSF is approximately radially symmetric.  As the
reconstructed extension of \nb\ is compatible with that expected for a
point source, the distribution of \g-rays from \nb\ is expected to be
radially symmetric.  Five circular spill-over test regions of a radius
of $0.07^{\circ}$ are placed at the same angular distance as between
\nb\ and \dorc, but rotated around \nb, as shown in
Fig.~\ref{fig:test_regions}, to estimate the \g-ray excess contributed
by N 157B in the \dorc\ on-region.

In total, $150\pm25$ \g-like excess (using the reflected
  background method \cite{BGmodels}) have been recorded in the test
regions, with an area 5 times larger than the on-region.  The average
\g-ray excess contributed by N 157B in the \dorc\ on-region is thus
estimated to be $(150\pm25)/5=30\pm5$.  Excluding the region of
highest or lowest excess does not significantly change this number,
demonstrating the robustness of this estimate with respect to the
location of the spill-over test regions.  The excess from \dorc\
before correction is $104\pm14$, implying a flux contamination of
$29\%\pm 6\%$.  The measured flux was correspondingly scaled down by a
factor $0.71\pm 0.06$.  As the reconstructed \g-ray spectral index of
the superbubble and the PWN are consistent within errors and the
energy dependence of the PSF is small in the energy range discussed
here, the correction factor is considered to be independent of
reconstructed \g-ray energy.

Another method to calculate the true excess from \dorc\ is to use the
spill-over test regions as background regions. In total, $183$ and
$544$ \g-like events (including true \g-rays and \g-like CRs) have
been recorded in the \dorc\ on-region and the test regions,
respectively. The excess from \dorc\ calculated in this way is
$74\pm14$, consistent with the value derived from the previous method.

\subsection{Fermi-LAT analysis}
\label{sec:fermi}
A search for GeV \g-rays from the sources detected by H.E.S.S. has
been carried out using data from the space-born Fermi-LAT
instrument. The LAT is a pair-conversion telescope, operating in the
energy range between $\sim20$\,MeV and $\sim300$\,GeV. A description
of the instrument- and mission-related details is given in
\cite{Fermi:Inst}. Based on the data from an 11 month observation
period, the detection of diffuse emission from the Large Magellanic
Cloud, best described by either a two-component Gaussian template or
an HII template, has been reported\cite{Abdo2010}. Emission from
individual point-like sources above this diffuse emission has not been
detected in this data-set.
 
To derive the \g-ray flux upper limits shown in Figure~4 in the main
article, more than 5 years of spacecraft data (2008-08-04 --
2013-10-21) have been analyzed using the Fermi Science Tools (FST)
package with version
v9r31p1\footnote{http://fermi.gsfc.nasa.gov/ssc/data/analysis/documentation/Cicerone/}. Only
events between $(10-300)$\,GeV have been considered, which results in
an excellent angular resolution of $<0.2^\circ$ and reduces the
contribution from the diffuse emission component. Events within a
$20^\circ$ by $20^\circ$ region and with zenith angles below
$100^\circ$, centered on RA \HMS{05}{17}{36.0} and Dec
\DMS{-69}{01}{48.0} (J2000) were analyzed with a binned likelihood
analysis using the P7Clean\_V6 instrument response functions.

Sources have been modeled according to the Fermi 2-year catalog
\cite{Fermi:2FGL}, and additional point-like sources at the positions
of \dorc, \nd, \nb, and \sn\ have been added. Note that significant
diffuse emission has been detected towards \nb \cite{Abdo2010}. As the
origin of this emission cannot unambiguously be attributed to one of
the H.E.S.S. sources, the narrow Gaussian component (G2 in
\cite{Abdo2010}) was removed from the model. The normalization of all
sources within $15^\circ$ as well as that of all diffuse sources has
been left free in the fit.

The \g-ray flux upper limits have been derived in three energy bands:
$(10-30)$\,GeV, $(30-100)$\,GeV, and $(100-300)$\,GeV, assuming a
power-law in energy with spectral indices as reconstructed for the
H.E.S.S. sources (see Table~1). In the case of \sn\, only flux upper
limits could be derived in the TeV regime, therefore a spectral index
of $\Gamma=1.8$, as motivated by \cite{Berezhko2011}, is assumed.

\subsection{SED modelling of \dorc\ and \nd}
The leptonic and hadronic model curves shown in Figure 3 in the main
paper have been obtained using a model for the time-dependent
injection and interaction of electrons and protons
(e.g. \cite{Abramowski2014}).

For \dorc\, electrons and protons are injected at a constant rate and
cool via synchrotron and Inverse Compton processes, and proton-proton
collisions, respectively. Given the rather low densities under
consideration, we ignore Bremsstrahlung and Coulomb losses. The
energy-dependent proton-proton cross-section is taken from
\cite{Kamae2006}. In a purely hadronic scenario, a total energy in
interacting protons of
$W_{\rm pp} = 0.7\times 10^{52}\,(n_{\rm
  H}/1\,\rm{cm}^{-3})^{-1}$\,erg
is required to explain the H.E.S.S. data (correcting for a 30\%
contamination from \nb, and assuming a cutoff energy of 100\,TeV and
proton spectral index $\Gamma_p=2.0$). This energy needs to be
increased to
$1.0\times 10^{52}\,(n_{\rm H}/1\,\rm{cm}^{-3})^{-1}$\,erg, if the
cut-off energy in the proton spectrum is reduced to 30\,TeV. When
assuming a different proton spectral index, the required energy in
interacting protons increases significantly to e.g.
$W_{\rm pp} = 25\times 10^{52}\,(n_{\rm H}/1\,\rm{cm}^{-3})^{-1}$ for
the $\Gamma_p=2.4$ case.

A source of uncertainty in the leptonic scenario is the different
radiation fields that contribute to the Inverse Compton process. One
component is the 40\,K radiation field from \dorc, which we assume to
have a constant energy density of 0.5\,eV\,cm$^{-3}$. The second
component originates from the 30 Doradus region. Depending on the
integration region of the Tarantula nebula, and the definition of the
background region, the radiation field energy density varies between
$\sim$0.5\,eV\,cm$^{-3}$ and $\sim$1.5\,eV\,cm$^{-3}$. Note that these
estimates assume that the Tarantula nebula and \dorc\ are at the same
projected distance. The derived magnetic fields for these radiation
fields vary between 10$\mu$G and 18$\mu$G. The energy in electrons is
$3.8\times10^{48}$\,erg to $2\times10^{49}$\,erg for the different
models. The X-ray data for \dorc\ are from \cite{Yamaguchi2009} and
Fermi-LAT limits on the GeV \g-ray flux have been derived as discussed
above.

The modelling of the \nd\ SED has been performed in a similar way to
the modelling of \dorc. Radio data are from \cite{Dickel1995}, whereas
for the upper limit on the X-ray synchrotron flux from \nd\ we
reanalyzed Chandra data, and based our estimates only on the
$(4.5-6.3)$\,keV band, which is poor in X-ray line emission (see
below). The SNR shell is interacting with a nearby molecular cloud,
which is also bright in infrared wavelengths. Given the complex
morphology and limited angular resolution of HESS, it is not clear
where the HESS emission is coming from. We therefore assume that all
the infrared emission (which has a total flux of $\sim$3\,Jy,
\cite{Tappe2006}) peaks at about 20\,$\mu$m ($\sim$145\,K temperature)
and is located in the shell, which has a radius of 11\,pc.  The actual
infrared radiation energy density experienced by the VHE electrons
depends on the locations of the VHE electrons, which could be the
forward, but perhaps also the reverse shock, with respect to the
location from which infrared emission is emitted.  If the VHE
electrons are located all around the forward shock, and the IR
emission emerges from within \nd\ all IR photons can in principle be
up scattered. Here we assume that about 50\% of the photons encounter
regions with VHE electrons. Under this assumption the leptonic
scenario for the combination of \g-ray flux and the X-ray synchrotron
flux upper limit requires a magnetic field of 20$\mu$G. Ignoring the
IR radiation fields, and thus only considering the Cosmic Microwave
Background Radiation (CMBR), the derived magnetic field is
15$\mu$G. The 20$\mu$G field is roughly a factor two lower than the
equipartition magnetic field in this object. Note, however, that a
detailed modelling of the radiation field is beyond the scope of this
paper. All input parameters for the non-thermal emission model are
summarized in Table~\ref{tab:modelling}.

\begin{table*}
\centering
\begin{tabular}{lccccccccc}
  \hline
  \hline
  Parameter & age & $\Gamma_{e,p}$ & $E_c$ & $B$ & $n_H$ & $W_e$ &
  $W_p$ & $T_1$, $T_2$ & $u_1$, $u_2$ \\
  & yrs & & TeV & $\mu$G & cm$^{-3}$ & $10^{49}$\,erg &
  $10^{49}$\,erg & K & eV\,cm$^{-3}$ \\ \hline
  \dorc & 6000 & 2.0 & 100 & 15 & 40 & 0.38 & 15 & 88, 40 & 1.5, 0.5\\
  \nd & 2350 & 2.0 & 30 & 20 & 40 & 1.0 & 25 & 145 & 1.0 \\
  \hline
  \hline
\end{tabular}
\caption{Input parameters used for the time-dependent modelling of
  \dorc\ and \nd. The electron/proton distribution is modelled as
  $dN/dE \propto E^{-\Gamma_{e,p}}\rm{exp}(-E/E_c)$. $T$ and $u$ are the
  temperatures and energy densities characterizing the infrared
  radiation fields. In \dorc\ two radiation field components are
  considered, while for \nd, only one component is considered. In all
  cases, the energy density of the CMBR ($1$~eV/cm$^{-3}$) is also
  taken into account for the Inverse Compton process.}
\label{tab:modelling} 
\end{table*}

\subsection{X-ray Analysis of \dorc\ and \nd}

The X-ray image of \dorc\ shown in Fig.~1 in the main article was
obtained from XMM-Newton observations (Observation IDs 113020201 and
104660301). Only the MOS1 and MOS2 detectors were used. The data
reduction was carried out using the XMM-Newton software of the
XMM-Newton Science Analysis Software (SAS,
http://xmm.esac.esa.int/sas/). Calibrated event lists were produced
for each exposure using the SAS emchain script. The periods affected
by soft proton flaring were excluded \cite{PrattArnaud2003}. To
generate images in the energy band $0.5 - 8$~keV, we created the
quiescent particle background (QPB) images, the count images and model
exposure maps for each observation and each instrument, using
mos-spectra and mos back scale. The combined mosaic map was finally
computed taking into account the different efficiencies of the two
instruments. The image has been smoothed with a Gaussian with a width
of $10\arcsec$.

The upper limit on the X-ray synchrotron emission of \nd\ in the
$4.5-6.3$\,keV continuum dominated band is based on the Chandra X-ray
observation of January 9, 2006 (ObsID 5532) processed with the CIAO
v4.5 software. We conservatively assumed that all the continuum is
X-ray synchrotron emission, whereas in reality a thermally dominated
origin for the continuum emission is more likely given that the
spectrum is rich in line emission.

\subsection{Radiation Fields in \nb}

The magnetic field estimates presented in the \nb\ section in the main
article sensitively depend on the measured \g-ray and X-ray spectra,
as well as the ratio between the energy densities of magnetic fields
and radiation fields. Depending on the location of \nb\ in the LMC
along the line of sight, different radiation fields can potentially
act as target for the IC scattering of high-energy electrons
accelerated in \nb. Besides the omnipresent CMBR, also the infrared
radiation from the 30~Doradus star-forming region and the OB
association LH~99 can potentially contribute (if not
dominate). Table~\ref{tab:rad_fields} summarizes the properties of all
radiation fields that potentially act as targets for the high-energy
electrons accelerated in \nb.

We study two extreme cases with maximum and minimum radiation field
energy densities to derive upper and lower limits on the magnetic
field in the PWN. In Scenario A it is assumed that \nb\ is related to
LH~99 and is located at the same projected distance as this OB
association, implying that the CMBR, as well as radiation associated
with 30~Doradus and LH~99, contribute to the target radiation
fields. In Scenario B, \nb\ is assumed to be unrelated to LH~99 and to
have large distances both to the OB association and to 30~Doradus so
that only the CMBR contributes.

Several authors argue that \nb\ is located in or behind the OB
association \cite{Chu1992, Wang1998, Chen2006}, which would support
Scenario A. In this case the dust emission of LH~99 would be the
dominant radiation field. The peak of the corresponding infrared
emission is centred on the HII region associated with the young stellar object 2MASS~J05375027$-$6911071, which is at a projected
distance of $\sim1\arcmin$ from \nb. As shown in \cite{Micelotta2009},
a two-component (cold dust and warm dust) modified black body model
best describes the spectral energy distribution of the infrared
emission from the OB association.
\begin{table}
\centering
\begin{tabular}{lcc}
  \hline
  \hline
Radiation Field & Temperature & Energy Density\\ 
& K & eV\,cm$^{-3}$ \\
\hline 
CMBR   & 2.7 & 0.26 \\
30~Doradus & 88 & 2.7 \\
LH~99 (cold dust) & 29 & 12.7 \\
LH~99 (warm dust) & 230 & 5.7 \\
\hline
  \hline
\end{tabular}
\caption{Radiation field energy densities and temperatures.}
\label{tab:rad_fields}
\end{table}
The LH~99 components have been modeled by the modified blackbody model
proposed in \cite{Micelotta2009}, whereas the CMBR and 30~Doradus
components are modeled as pure blackbody emitters.  Since \nb\ is
potentially inside the 2MASS~J05375027$-$6911071 HII region, the
derived energy densities were furthermore scaled up by a factor of
1.5, as would be expected for an object at the edge of a homogeneously
emitting sphere \cite{Atoyan1996}.
\begin{figure} 
\centering
 \includegraphics[width=\columnwidth,draft=false]{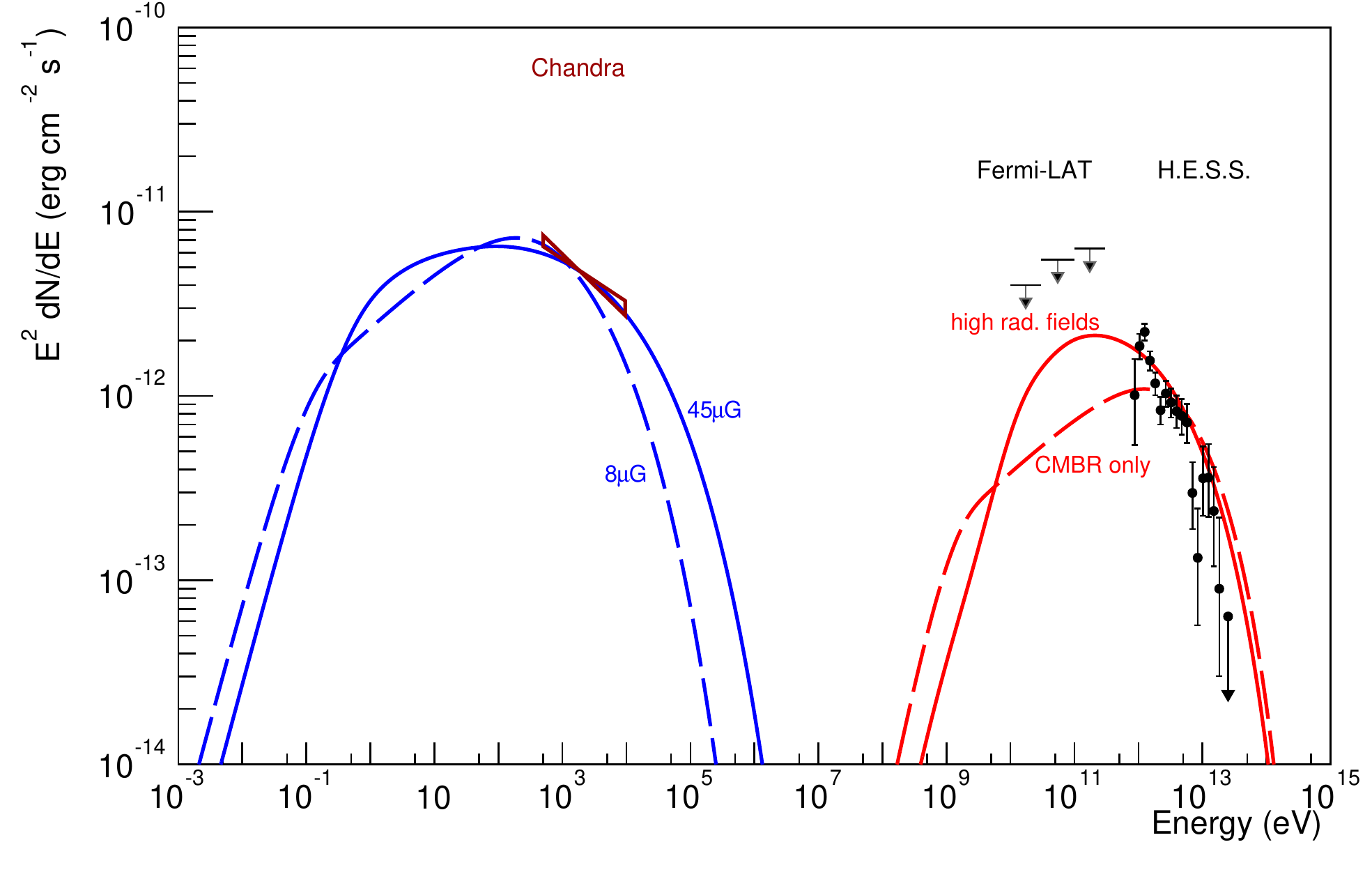}
   \caption{Spectral energy distribution of \nb. Also shown are the
     two scenarios as discussed in the text for maximum (minimum)
     radiation fields as straight (dashed) lines. A maximum (minimum)
     magnetic field of 45\,$\mu$G (8\,$\mu$G) is required to still fit
     the X-ray and \g-ray data, respectively. For the high-radiation
     field model a power-law index of injected electrons $\Gamma_{e,1}
     = 2.0$ and energy in electrons $\epsilon_{e,1} = 0.07$ of the
     current pulsar spin-down power is assumed. For the low-radiation
     field model, $\Gamma_{e,2} = 2.35$ and $\epsilon_{e,2} = 0.45$
     are assumed. In both scenarios, the cut-off energy is fixed to
     $E_{\rm e,c} = 100$\,TeV.}
\label{fig:rad_fields}
\end{figure}

Figure~\ref{fig:rad_fields} shows the spectral energy distribution of
\nb\ from radio wavelength to \g-ray energies and the broadband
synchrotron and IC emission expected for the time-independent
injection of relativistic electrons ($400\,\rm{GeV} \le E_e \le
100$\,TeV) over the lifetime of \psr. In Scenario A an injection of
electrons at a constant rate of 7\% of the current spin-down power of
\psr\ is required to explain the SED, i.e. $\dot{W} /
\dot{E}_{\rm{now}} = 0.07$. In this case, a magnetic field of
45\,$\mu$G is required to fit the data. Since the cooling time of
electrons that produce the observed X-ray and TeV emission would be
very short (i.e. $\lesssim 300$ years for $E_e > 20$\,TeV), the
spin-down power of the pulsar should not have changed significantly on
the scale of the electron cooling time. Consequently, this magnetic
field estimate is rather insensitive to the power-law index or cut-off
energy of the injected electrons, as well as to the age or braking
index of the pulsar. In contrast, in Scenario B, a much smaller
magnetic field strength is required to explain the data. At the same
time, the power injected into the nebula needs to be increased to
$\sim$45\%. The smallest field strength still compatible with the
X-ray and TeV \g-ray data is 8\,$\mu$G, although the spectral shape at
keV energies is poorly reproduced in the low-field case. This mismatch
could be overcome if a distribution of magnetic field values is
considered, instead of a single-magnetic field. The power injected
into the nebula of \nb\ in this scenario is comparable to the Crab,
however, the magnetic pressure would be much lower (a factor
$\sim$240). In this scenario, electron cooling times are much longer
(i.e. $\lesssim 8400$ years for $E_e > 20$\,TeV) and the model is less
reliable.

Note that the modelling of the \g-ray spectrum is insensitive to the
exact choice of the low-energy cut-off in the electron spectrum
(400\,GeV in this case) as H.E.S.S. spectral points only start at
$\sim$800\,GeV. Even higher-energy electrons are responsible for the
observed X-ray emission. The required fraction of pulsar spin down
injected into the nebula, however, does change with the minimum
electron energy.

\subsection{On the flux of \nb}

The differential flux of \nb\ of $\Phi(1\,\rm{TeV}) = (1.3 \pm 0.1)
\times \rm{10^{-12}\,cm^{-2}\,s^{-1}\,TeV^{-1})}$ increased by 64\%
compared to the previously published value \cite{2012A&A...545L...2H}.
A reanalysis of the data set which was used in the previous
publication yields spectral results compatible to the values reported
here. Thus, the change of the reported flux is not intrinsic to the
source, the \g-ray emission of \nb\ should be considered as being
constant.

The results presented here and in the previous publication
\cite{2012A&A...545L...2H} use an image-fitting analysis
\cite{Mathieu} which provides an improved angular resolution and a
higher \g-ray efficiency. But this reconstruction is also much more
susceptible to imperfections in the detailed modelling of the
instrument response than the classical Hillas-based analysis. One such
imperfection was a misaligned camera which is now corrected in the
analysis. In the initial publication \cite{2012A&A...545L...2H} this
correction was not yet taken into account. The misaligned camera
results not only in an underestimated extend of the point-spread
function but also in a global shift of the \textit{ShowerGoodness},
the main event selection parameter \cite{Mathieu}. The global shift of
the \textit{ShowerGoodness} parameter leads to a misclassification of
actual \g-rays as background and thus an underestimation of the true
\g-ray flux.

This systematic error is relevant only for a small number of
publications, which utilise the image-fitting analysis
\cite{Mathieu}. A re-analysis of affected sources is currently
underway and errata will be published in due time.

\end{document}